# Weak localization and electron-electron interactions in Indium-doped ZnO nanowires

*Richard S. Thompson,* [*,†] *Dongdong Li,* [‡,†] *Christopher M. Witte,*[†] *and Jia G. Lu*[*,†]

[†] Department of Physics and Department of Electrical Engineering, University of Southern California, Los Angeles, California 90089-0484

[‡]School of Materials Science and Engineering, Shanghai Jiao Tong University, Shanghai 200240 (China)

[*] To whom correspondence should be addressed. Email: rsthom@usc.edu and jia.grace.lu@usc.edu.

ABSTRACT

Single crystal ZnO nanowires doped with indium are synthesized via the laser-assisted chemical vapor deposition method. The conductivity of the nanowires is measured at low temperatures in magnetic fields both perpendicular and parallel to the wire axes. A quantitative fit of our data is obtained, consistent with the theory of a quasi-one-dimensional metallic system with quantum corrections due to weak localization and electron-electron interactions. The anisotropy of the magneto-conductivity agrees with theory. The two quantum corrections are of approximately equal magnitude with respective temperature dependences of $T^{-1/3}$ and $T^{-1/2}$. The alternative model of quasi-two-dimensional surface conductivity is excluded by the absence of oscillations in the magneto-conductivity in parallel magnetic fields.

With the advent of nanotechnology, semiconducting nanowires have attracted extensive interest as the building blocks for nanoelectronic devices [1-4]. ZnO and its doped form have been recognized as a promising material system with versatile functionalities based on its remarkable electric, piezoelectric, optoelectric and magnetic properties. Undoped ZnO exhibits *n*-type semiconducting behavior, originating from native defects, mainly of Zn interstitials, oxygen vacancies or hydrogen interstitials. It has been found that doping impurities, such as Mg,[5] Ga,[6] In,[7-9] P, [10] Sn, [7] and Al, [11] can significantly enhance the electrical conductivity, leading to promising applications in nanoelectronic devices. On the other hand, nanowires' quasi-one-dimensional (Q1D) structure provides a natural paradigm for studying the fundamental physical principles. In particular, the transport conduction theories, such as hopping [4, 12] and weak localization, [13, 14] have been reported in both as-grown and doped ZnO nanowires. In addition, electron localization in various nanomaterials, such as Si,[15] $SnO_2$,[16] InAs[17] based nanowires, has been investigated at low temperatures. Although increasing effort has been devoted to this field, an in-depth comprehensive study of the electrical transport is still missing and the underlying mechanisms remain largely unclear. [18] With the aim to understand the microscopic mechanisms, we have performed experiments on the temperature and magnetic field dependent conductance of In-doped ZnO (In:ZnO) nanowires. Combined with theoretical modeling, we find an excellent quantitative fit of our data in the temperature range between 4.2 and 10 K, by treating In:ZnO as a disordered metallic system with corrections due to weak localization and electron-electron interaction effects in such Q1D structures.

The In:ZnO nanowires are synthesized by the laser-assisted chemical vapor deposition (CVD) method. Pure zinc powder is placed in the center of the furnace, where the temperature is elevated to 650 °C to generate zinc vapor in argon (Ar) under 1 atmosphere pressure. Oxygen (23 ppm) diluted by Ar is kept flowing at a rate of 220 sccm. Pure indium powder is simultaneously ablated from upstream under Nd:YAG (yttrium aluminum garnet) laser pulses. In-doped ZnO nanowires are formed on tin-coated (5 nm) silicon substrates via the catalytic vapor-liquid-solid process.[19] The energy dispersive X-ray (EDX) spectrum (as displayed in the lower left inset of Fig. 1) shows that the indium to zinc atomic ratio is around 3%. Selected area electron diffraction patterns (upper right inset of Fig. 1) and high resolution transmission electron microscopy (not shown in this paper) reveal the single crystalline structure with the (0001) growth direction, indicating that the In atoms are well incorporated into the wurtzite crystal lattice. Because of the mismatch of the atomic radii between In and Zn some defects are present.

The as-synthesized In:ZnO nanowires are suspended in isopropyl alcohol and then deposited onto degenerately doped Si substrates that are capped with a 500 nm $SiO_2$ layer. An individual nanowire is located by scanning electron microscopy (SEM), followed by four-electrode patterning via e-beam lithography. Post–thermal annealing at 300 ºC for 10 min in vacuum is utilized to further improve the contacts. The current-voltage (I-V) and current-gate voltage (I-$V_g$) curves taken at 4.2 K manifest very weak gate dependence. A plot of the resistance versus gate voltage $V_g$ is shown in Fig. 1 for two measurements. The oscillations persist when a magnetic field is applied but decrease in amplitude when the temperature is raised. Similar gate oscillations have been reported in InAs nanowires[17], attributed to the quantization of the momentum of the electrons due to the finite size of the nanowires. However, since the mean free path is found to be much shorter than the radius in our samples, the reason for the oscillations is uncertain and needs to be systematically studied.

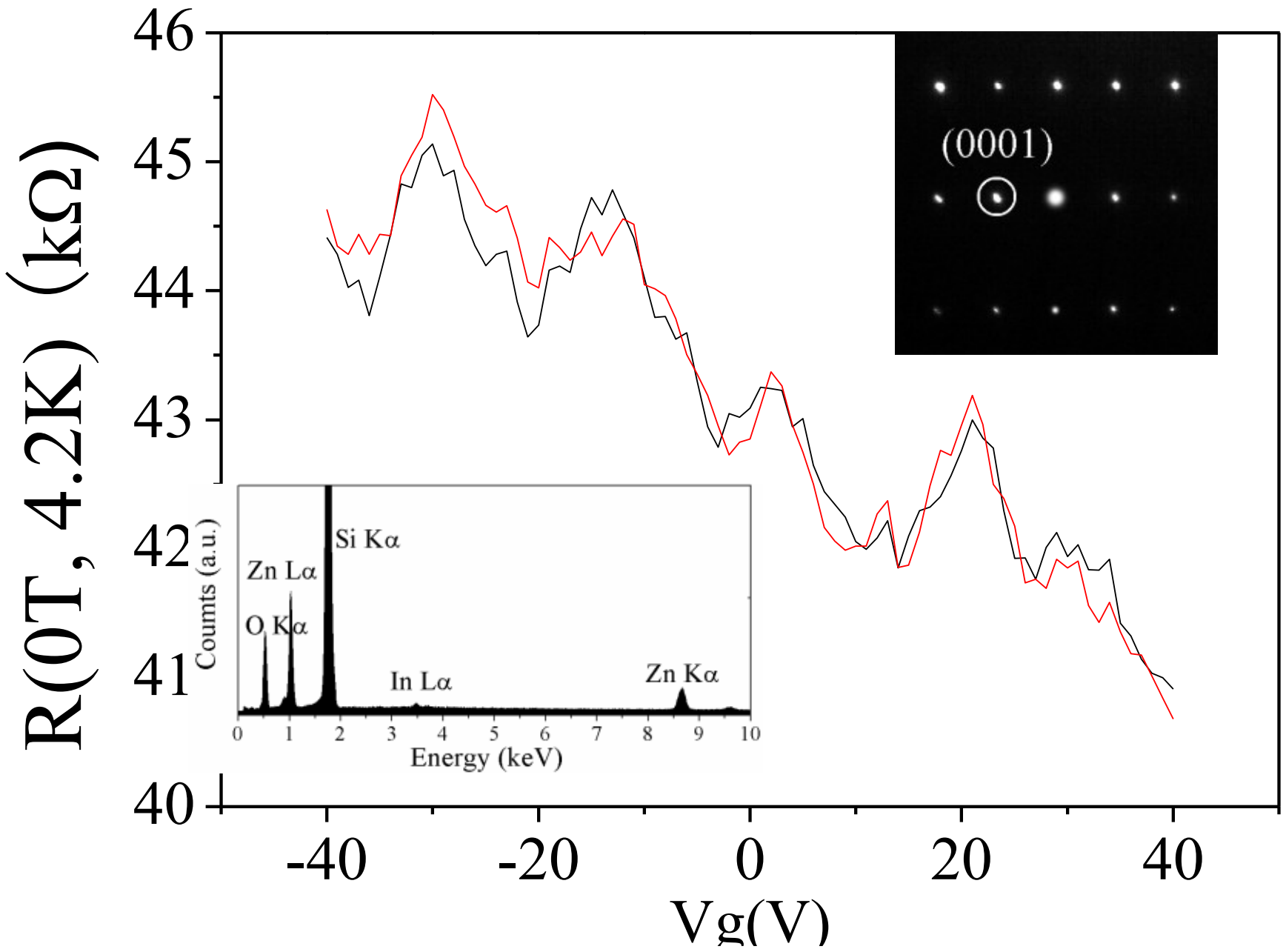

Figure 1  Resistance of a single nanowire versus gate voltage $V_g$ at temperature 4.2 K in zero magnetic field (two different runs are shown).  Lower left inset:  the energy dispersive X-ray (EDX) spectrum of In:ZnO nanowires.  Upper right inset: selected area electron diffraction pattern of an individual In:ZnO nanowire.

The SEM image in the upper left inset of Fig. 2 displays the configuration of a single nanowire (~40 nm in diameter) in contact with Ti/Au electrodes with 2 μm inter-electrode spacing for resistance measurements. The resistivity of the In-doped nanowire measured by the 4-probe technique at 4.2 K is 27 μΩ m, which is more than two orders of magnitude smaller than that of pure ZnO nanowires.[4]  The nanowires are in the heavy-doping regime and show degenerate metallic behavior at low temperatures with quantum corrections.  There are two important quantum corrections: weak localization and electron-electron interactions.  The key experiments to justify this interpretation are magneto-resistance measurements because they allow the derivation of the inelastic scattering length.  The magneto-resistance is mostly due to a reduction in the weak localization effect, while the two corrections have comparable temperature dependence.  Using this data, the temperature dependence of the quantum corrections is calculated, showing that these corrections account for the temperature dependence of the total conductivity.  This means that the basic metallic Drude conductivity is constant in this temperature range.

The magneto-resistance (MR) is measured with the magnetic field ranging from -2.75 to 2.75 T in the temperature range from 4.2 to 10 K. Fig. 2 plots the MR data with the magnetic field perpendicular to the nanowire axis, and the upper right inset shows the data with the magnetic field parallel to the wire axis. As explained in the following analysis section, the negative MR is a consequence of the suppression of quantum interference effects in the weak localization theory.[20, 21] It is important to notice that there are no oscillations observed in the MR on the scale of 1 T for the parallel field orientation.

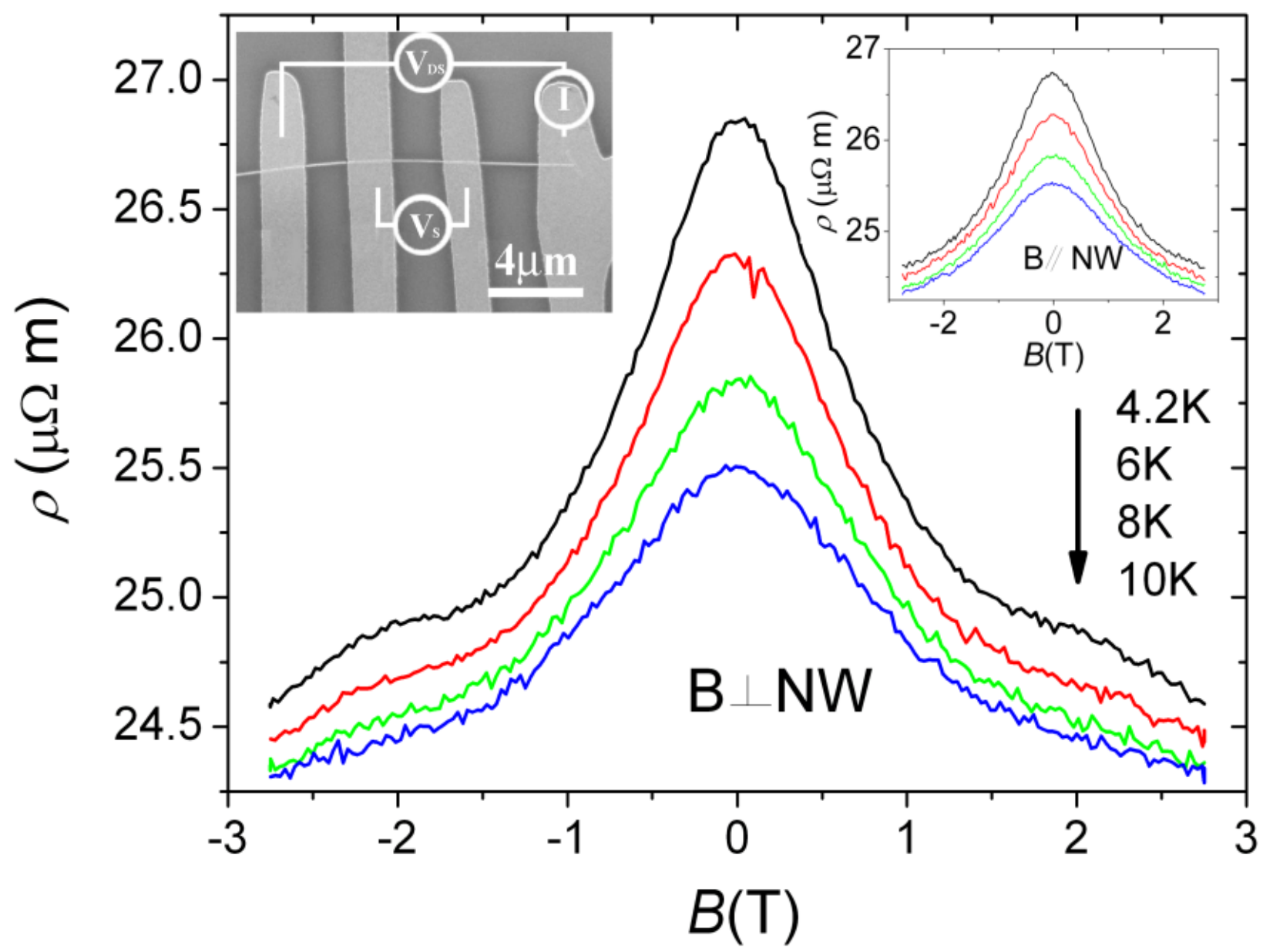

Figure. 2 Magneto-resistivity (MR) curves obtained with the magnetic field applied perpendicular to a nanowire axis at different temperatures ranging from 4.2 to 10 K. Upper left inset: SEM image of a single nanowire in contact with four electrodes. Upper right inset: MR plots with the magnetic field parallel to the nanowire.

Figure 3 shows the magnetic field dependence of the conductivity for small perpendicular field from 0 to 0.5 T in the temperature range from 4.2 to 10 K. The quadratic slope of the $\Delta\sigma - B^2$ curves for the magnetic field perpendicular to the wire axis is found to be about twice as large as for that in the parallel case (Fig. 3 inset). We will show below that this behavior is in agreement with weak-localization theory. To our knowledge, this is the first experimental verification of this predicted asymmetry, since the previous experiments focused only on the perpendicular field orientation.

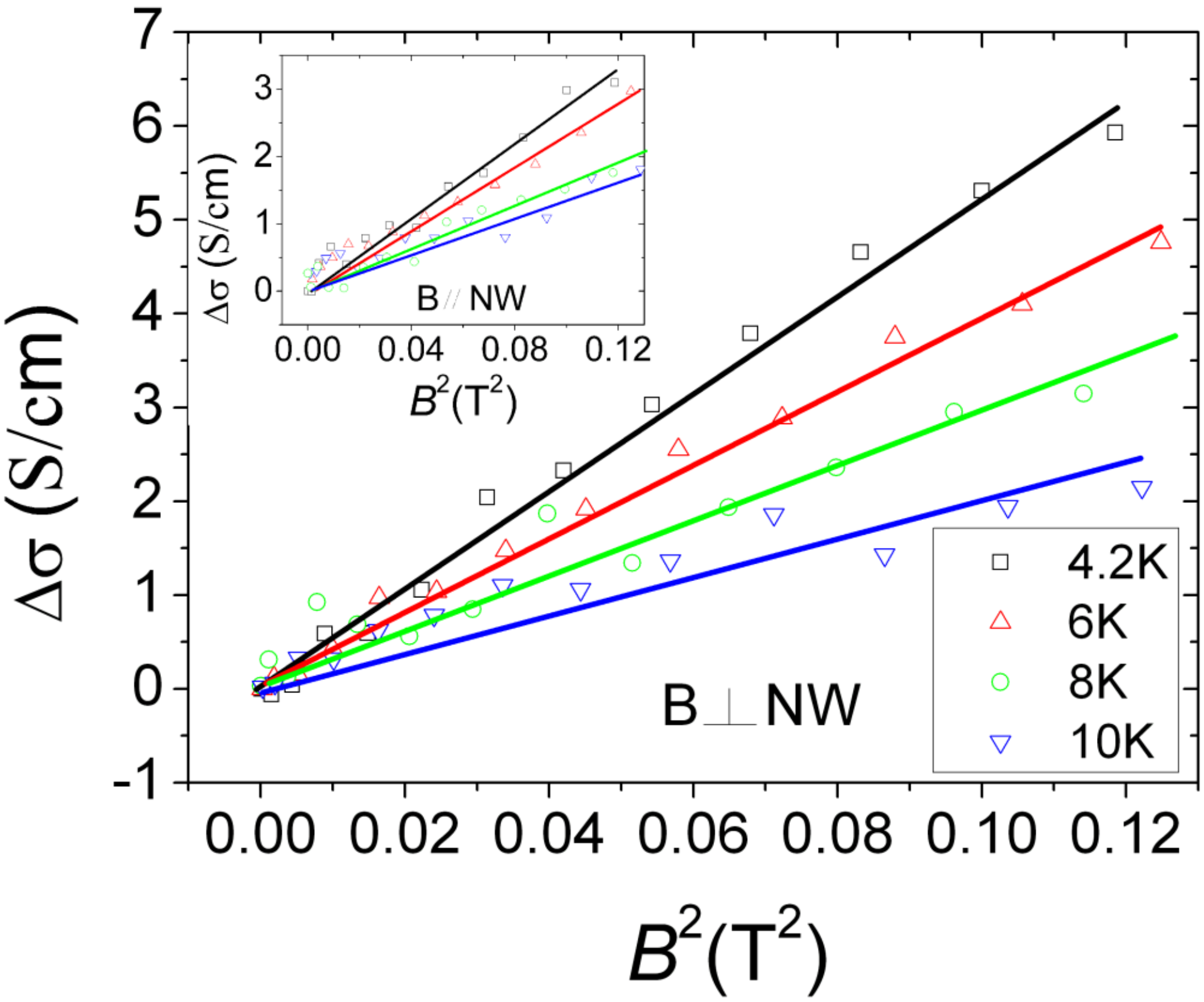

Figure 3. Magnetic field dependence of $\Delta\sigma = \sigma(B,T) - \sigma(0,T)$ in the small field range with the nanowire perpendicular to the field (for T = 4.2, 6, 8, and 10 K, respectively). The inset shows the magneto-conductivity for parallel fields. The quadratic slope for perpendicular field is found to be about twice as large as for parallel field.

Figure 4 plots the conductivity in small magnetic fields ranging from 0 to 0.5 T as a function of temperature between 4.2 and 22 K, showing a power law behavior of $T^{-1/2}$.

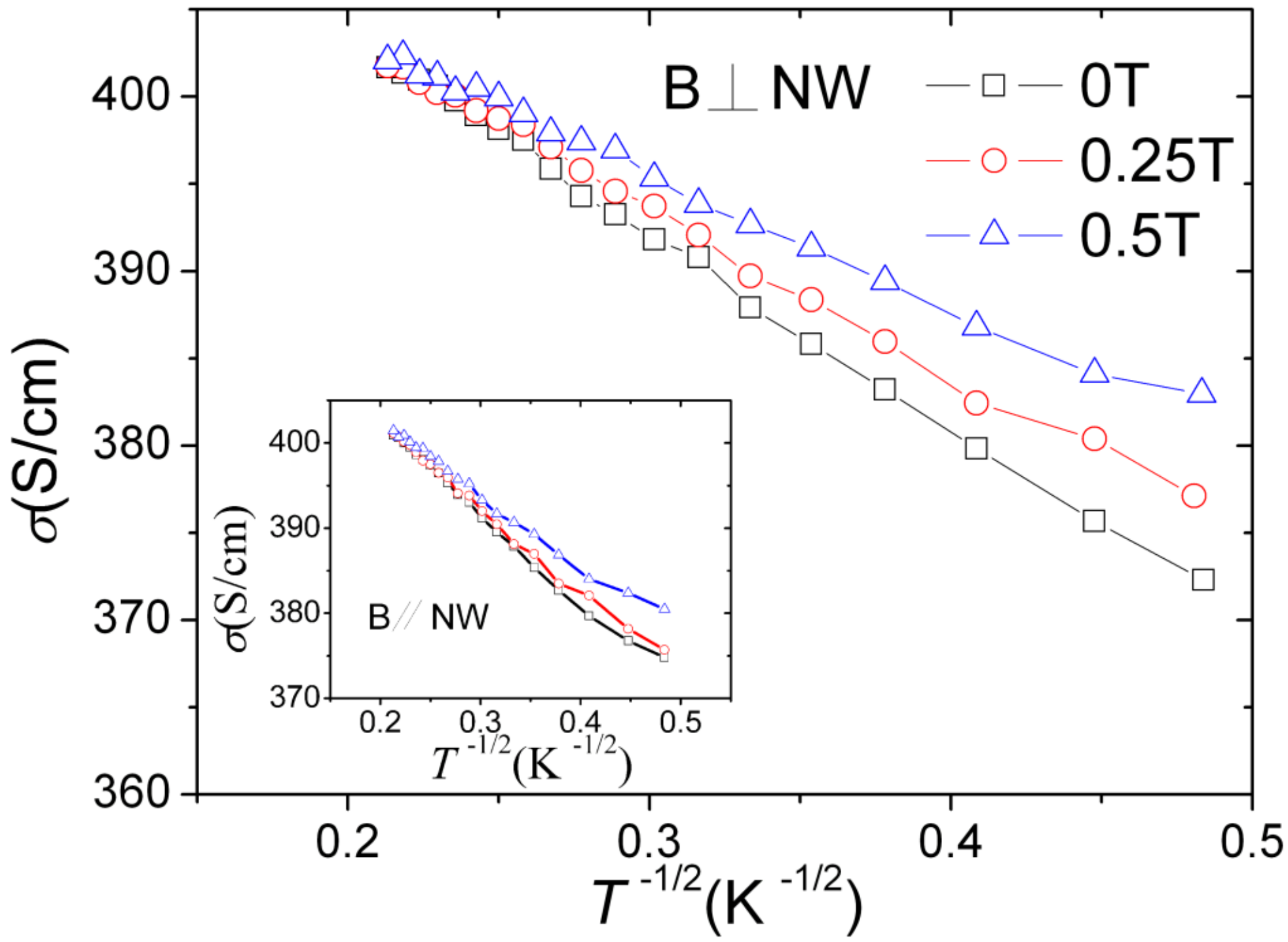

Figure 4. Temperature dependence of conductivity for the field perpendicular to the wire axis with values of B = 0, 0.25, and 0.5 T, respectively. The inset shows the data for parallel fields. In the temperature range from 4.2 to 22 K the conductivity decreases as a function of temperature with an approximate power law of $T^{-1/2}$.

Due to the large surface-to-volume ratio, it has been a question whether the conduction is dominated by the surface layer. It was proposed in refs. 13 and 14 that the temperature dependence of the conductivity at low temperatures of similar nanowires was proportional to ln($T$), and therefore it was suggested that these temperature corrections might be due to two-dimensional weak localization in a surface layer of their wires. However, our experiments on the magneto-conductivity in parallel fields rule out this interpretation. According to the experiments of Sharvin and Sharvin [22] on hollow metal cylinders, if the phase relaxation length is large enough relative to the wire radius, then one would expect to see magneto-conductivity oscillations in parallel field with a period of the flux quantum for pairs $BS = h/2e$, where $S$ is the wire cross-sectional area. For our wires of radius 20 *nm*, the corresponding field is B = 1.6 T. As shown in the inset to Figure 2, there is significant magneto-conductivity, but no such oscillation effect is observed. Therefore, we instead analyze our results based on solid metallic behavior in this temperature range using the theory of quantum corrections to the conductivity due to weak localization and the electron-electron interaction. [23, 24] In our case the inelastic diffusion length is longer than the wire radius, which sets the system in the quasi-one-dimensional regime.

The normal-state characteristic parameters for our samples are calculated in the following manner using three input parameters. From the data in Figure 4, we assume that the zero-field conductivity in this low temperature range is given by the sum of two terms $\sigma = \sigma_0 + \Delta\sigma$, where $\sigma_0 = ne^2\tau / m^*$ is the temperature-independent Drude conductivity and $\Delta\sigma$ is the temperature-dependent quantum correction. By extrapolating the linear behavior of our data to $T^{-1/2} = 0$ we get our first parameter $\sigma_0 = 4.25\times10^4\ S/m$. Our second parameter, the carrier concentration *n*, is obtained from the dependence of the resistance on the gate voltage shown in Fig. 1. For a typical doped semiconductor one obtains the carrier density *n* by assuming that the mobility $\mu = e\tau / m^*$ is constant as the gate voltage is changed. However, this assumption is not valid when the carriers form a degenerate Fermi gas. The elastic scattering time $\tau$ in a metal is given by the formula[25] $\frac{1}{\tau} = \frac{n_i m^* p_F}{(2\pi)^3}\int |u(\theta)|^2\, d\Omega$, where $n_i$ is the impurity concentration, *u* is the Fourier transform of the impurity scattering potential, $\theta$ is the scattering angle, and $\Omega$ is the solid angle. The Fermi momentum $p_F$ is proportional to the carrier density to the power 1/3. Therefore, the conductivity $\sigma$ is proportional to $n^{2/3}$, and $\frac{d\sigma}{\sigma} = -\frac{dR}{R} = \frac{2}{3}\frac{dN}{N}$, where *N* is the total number of carriers in the wire. The change in *N* is related to the change in the gate voltage by the capacitance $C = e\frac{dN}{dV_g}$ of the wire with respect to the gate plate. The usual formula $C/L = 2\pi\varepsilon_0 / \ln(2h/r)$ for the capacitance per length *L* of an infinitely long thin wire of radius *r* located a distance *h* above a conducting plane is obtained using the image method. When immersed in a material with relative dielectric constant $\varepsilon_r$, which for $SiO_2$ has the value of 3.9, the formula for *C* is multiplied by $\varepsilon_r$. However, in our case, the $SiO_2$ only fills the region between the wire and the plate and not the region on the other side of the wire, which is vacuum. In this case a reasonable approximation would be to multiply the above formula for *C/L* by the average dielectric constant (3.9+1)/2=2.45. Actually, it is possible to solve exactly the problem of an infinitely long thin wire above a conducting plane with $SiO_2$ between the wire and the plane by taking the Fourier transform in the direction parallel to the plane and solving the resulting one-dimensional equation in the direction perpendicular to the plane. The result is that the effective dielectric constant between the wire and the plate is 2.66. Consequently $C/L = 38\, pF/m$ and $C = 76\, aF$ for our length $L = 2\,\mu m$. Then from Fig. 1, we use $dR/dV_g = 45\Omega/V$ and $R = 4.3\times10^4\Omega$ and find that $N = 3.0\times10^5$ and $n = 1.2\times10^{26} m^{-3}$. Of course, the charges induced by the gate voltage are confined to a narrow surface layer by Thomas-Fermi screening, but since the surface conductivity is in parallel with the bulk conductivity, the actual distribution of charge should not be important. Finally, the third parameter is the effective electron mass, which is taken from the literature as the conduction-band mass $m^* = 0.29 m_e$.[26]

The additional characteristic parameters for our samples are calculated as follows by using these three parameters and the free-electron model. The Fermi wave number is $k_F = (3\pi^2 n)^{1/3} = 1.5\times 10^9 m^{-1}$, the Fermi momentum is $p_F = \hbar k_F = 1.6\times 10^{-25} kg\, m/s$, and the Fermi velocity is $v_F = p_F / m^* = 6.1\times 10^5 m/s$. Thus the Fermi energy is $E_F = 0.31 eV$, which is much larger than the thermal energy at 10 K, leading to a degenerate Fermi gas, as mentioned above. The elastic scattering time is calculated to be $\tau = \frac{m^* \sigma}{ne^2} = 3.6\times 10^{-15} s$. The mean free path is $\ell = v_F \tau = 2.2 nm$. Notice that the product $k_F \ell = 3.4$, setting the system in the metallic regime. Finally, the diffusion constant is determined to be $D = v_F \ell / 3 = 4.5\times 10^{-4} m^2/s$.

Next we consider the quantum corrections to the conductivity, starting with weak localization. The standard formula for the conductivity correction due to weak localization for a wire with a square cross section of width $a$ in the absence of a magnetic field, magnetic impurities, and spin-orbit scattering is $\Delta\sigma = -\frac{e^2}{\pi\hbar S}\sum_{m,n=0}^{a/(\pi\ell)} \frac{1}{\sqrt{(m^2+n^2)(\pi/a)^2 + (L_\varphi)^{-2}}}$, where $S$ denotes the cross-sectional area of the sample and $L_\varphi$ is the inelastic diffusion length $L_\varphi = \sqrt{D\tau_\varphi}$, with diffusion constant $D$ and the inelastic lifetime $\tau_\varphi$. In the one-dimensional regime where $a << \pi L_\varphi$, the first term in the sum dominates, giving $\Delta\sigma = -\frac{e^2}{\pi\hbar S} L_\varphi$. In the three-dimensional (3D) regime where $a >> \pi L_\varphi$, the sum is replaced by an integral yielding $\Delta\sigma = -\frac{e^2}{2\pi^2\hbar}\left(\frac{1}{\ell} - \frac{1}{L_\varphi}\right)$, which depends on a temperature-independent cut-off length $\ell$ on the order of the electron mean free path. In the intermediate range where $a \sim \pi L_\varphi$, we make a rough approximation of replacing the sum by the first term plus the integral, namely by adding the one-dimensional and three-dimensional equations.

The corrections to these equations due to a weak magnetic field have been calculated using first order perturbation theory by Altshuler and Aronov.[27] In the above formulas, $(L_\varphi)^{-2}$ must be replaced by $(L_{\varphi 0})^{-2} + (L_{\varphi B})^{-2} = (L_{\varphi 0})^{-2} + <(2eA/\hbar)^2>$, where **A** is the vector potential in an appropriate gauge and <> denotes the average over the sample. Their calculation is done for a circular sample in a parallel field and for a square sample in a perpendicular field in order to easily satisfy the required boundary conditions. The results are that $< A^2 >= \frac{B^2 S}{12}$ for the perpendicular field and $< A^2 >= \frac{B^2 S}{8\pi}$ for the parallel field. The perpendicular field is predicted to be twice as effective as the parallel field since $8\pi \approx 25$, which concurs with our measurements. The perturbation theory is valid when $\frac{2e}{\hbar} BS < 1$. In fact, for the perpendicular field orientation, there is another correction of order $B^2$ in second order perturbation theory due to the coupling of the vector potential with the component of the fluctuation momentum along the wire axis. This correction multiplies the one-dimensional expression for $\Delta\sigma$ by $\left[1 - \frac{1}{30}\left(\frac{2e}{\hbar} BS\right)^2\right]^{-1/2}$, which clearly indicates the limitation on the validity of perturbation theory mentioned above. We find that the three-dimensional corrections to the magneto-conductivity for our samples are very small, thus we have used the simple Q1D formula $\Delta\sigma(T,B) - \Delta\sigma(T,0) = \frac{e^2}{\pi\hbar S} \frac{(L_{\varphi 0})^3}{2(L_{\varphi B})^2} = \frac{e^2}{\pi\hbar}\left(\frac{2e}{\hbar}\right)^2 \frac{B^2}{24}(L_{\varphi 0})^3$ to calculate $L_{\varphi 0}$ from the initial slope of the quadratic magneto-conductance in perpendicular fields (from Fig. 3). (Note that the conductivity corrections due to the magnetic-field dependence of the electron-electron interactions are also negligible.) From the magneto-resistance data at temperature 4.2 K, we calculate $L_{\varphi 0} = 56.3\,nm$. $L_{\varphi 0}$ decreases smoothly to $L_{\varphi 0} = 40.7\,nm$ at 10 K. If one assumes that $L_{\varphi 0}$ is proportional to $T^{-p/2}$, the best fit to our data gives an exponent $p = 0.73$. The corresponding values of the phase relaxation time $\tau_{\varphi 0} = (L_{\varphi 0})^2 / D$ are $\tau_{\varphi 0} = 7.0 \times 10^{-12} s$ at 4.2 K and $\tau_{\varphi 0} = 3.7 \times 10^{-12} s$ at 10 K.

At low temperatures the main contribution to phase relaxation in one dimension comes from electron-electron interactions with small energy transfer (compared with $k_B T$), also called the Nyquist noise,[28] and the value of the exponent $p$ is predicted to be $p = 2/3$, in accord with our measurements. The theoretical formula for this Nyquist phase relaxation time is $\tau_N = \left[\frac{\hbar^4 \sigma^2 S^2}{2e^4 (k_B T)^2 D}\right]^{1/3}$. This formula gives $\tau_N = 5.6 \times 10^{-12} s$ at 4.2 K and $\tau_N = 3.1 \times 10^{-12} s$ at 10 K. The theoretical values of $\tau_N$ are a factor of 0.8 smaller than our above experimental numbers for $\tau_{\varphi 0}$ .

However, according to ref. 28 the standard theory given above does not apply when the phase relaxation is dominated by electron-electron interactions with small energy transfer due to a break-down of the quasiparticle description of the Fermi liquid. The revised expression in the one-dimensional case is $\Delta\sigma(T,B) = -\frac{e^2 L_N}{\pi\hbar S}\left[1.94 - \sqrt{32}\,\frac{(L_N)^2}{2(L_{\varphi B})^2}\right]$ when expanded for small magnetic fields.[30] This formula is the same as the standard theory except for the replacement of $L_{\varphi 0}$ by $L_N = \sqrt{D\tau_N}$ and the two new coefficients: 1.94 and $\sqrt{32}$. The result of using this corrected formula is that our values of $L_N$ are reduced by a factor of $(32)^{1/6} = 1.78$ from the values of $L_{\varphi 0}$ stated above, and our experimental values of $\tau_N$ are reduced by a factor of $(32)^{1/3} = 3.17$ from the values of $\tau_{\varphi 0}$ given above. Now our experimental values are smaller than the theoretical results by a factor of 0.4. It should be noted that in ref. 29, in spite of the good agreement of the temperature exponent of $\tau_N$ over several decades, there are similar difficulties in fitting the prefactor.

The revised formula gives almost exactly the same zero-field temperature dependence of $\Delta\sigma$ because the two correction factors almost exactly compensate each other, *i.e*. $1.94/1.78 = 1.09$. The magnitude of the one-dimensional conductivity correction due to weak localization remains the same, and the temperature dependence remains proportional to $T^{-1/3}$. The one-dimensional weak localization theory gives a conductivity difference between 10 and 4.2K of $\frac{e^2}{\pi\hbar S}\left[L_\varphi(4.2K) - L_\varphi(10K)\right] = 962\,(\Omega m)^{-1}$. The three-dimensional weak localization theory gives a conductivity difference of $\frac{e^2}{2\pi^2\hbar}\left[\frac{1}{L_\varphi(10K)} - \frac{1}{L_\varphi(4.2K)}\right] = 84\,(\Omega m)^{-1}$. Thus, the Q1D effect is much more significant than the 3D correction.

Furthermore, there are still the electron-electron interaction effects to consider. The contributions to the conductivity due to the electron-electron interaction (also called the Coulomb anomaly) are similar to those due to weak localization but with the length $L_\varphi$ replaced by the thermal diffusion length $L_T = \sqrt{\frac{\hbar D}{kT}}$, and the coefficients are slightly modified. The exponent of the temperature dependence $L_T$ is -1/2, the same as we have used in Fig. 4. In the Q1D regime, $\Delta\sigma_{ee} = -\left[0.39\left(4 - \frac{3\tilde{F}}{2}\right)\right]\frac{e^2}{\pi\hbar S}L_T$, while in the 3D regime $\Delta\sigma_{ee} = -\left[0.46\left(\frac{4}{3} - \frac{3\tilde{F}}{2}\right)\right]\frac{e^2}{2\pi^2\hbar}\left(\frac{1}{\ell} - \frac{1}{L_T}\right)$. From our material parameters we calculate the correction factor to be $\tilde{F} = 0.6$. Using our estimated value for the diffusion constant $D = 4.5\times10^{-4}\,m^2/s$, we calculate $L_T = 28.6\,nm$ at 4.2 K and $L_T = 18.5\,nm$ at 10 K. Then the calculated conductivity differences between 10 and 4.2 K due to the electron-electron interaction are $751(\Omega m)^{-1}$ (Q1D term) and $47\,(\Omega m)^{-1}$ (3D contribution).

Altogether considering the 3D and Q1D weak localization and electron-electron interaction corrections, the sum of the four calculated contributions to $\Delta\sigma$ between 10 and 4.2 K is $1844(\Omega m)^{-1}$. The magnitudes of the weak localization and the electron-electron corrections are approximately equal. Experimentally our conductivity at 4.2 K is $3.723\times10^{4}(\Omega m)^{-1}$ and at 10K is $3.908\times10^{4}(\Omega m)^{-1}$, so the difference is $1850(\Omega m)^{-1}$. Therefore, we believe that the combined theory is in excellent agreement with our data.

In conclusion, we have presented low temperature magneto-conductance measurements on ZnO nanowires highly doped with In. The temperature and magnetic field dependence of the conductivity of our wires is in good agreement with the theory of quasi-one-dimensional weak localization and electron-electron interactions in a disordered metallic system. In particular, the quadratic magneto-conductance slope for the magnetic field perpendicular to the wire axis is found to be about twice as large as for that in parallel case. To our knowledge, this is the first experimental verification of this predicted asymmetry. In addition, the absence of oscillations in the magneto-conductance curves for magnetic fields parallel to the nanowires excludes the previous suggestions and indicates that the conductance of these nanowires is a bulk, rather than a surface effect.

**Acknowledgment.** The authors are grateful to Professor Gerd Bergmann for countless insightful discussions and suggestions. We also thank Mr. Zuwei Liu, Mr. Chiachi Chang, and Mrs. Yichen Chang for their helpful assistance in low temperature measurements. This work is supported by NSF grant ECS-0306735. Dongdong Li is also grateful for the support of the China Scholarship Council.